\documentclass[preprint,12pt]{elsarticle}

\usepackage{amssymb}
\usepackage{graphicx,float}
\usepackage{dcolumn}
\usepackage{bm}
\usepackage{amsmath}
\usepackage[latin1]{inputenc}
\usepackage{psfrag}
\usepackage{epsfig}

\usepackage[nodots,nocompress]{numcompress}

\biboptions{super}

\journal{Annals of Physics}

\begin{document}

\begin{frontmatter}



\title{Thermofield qubits, generalized expectations and quantum information
protocols}


\author{T. Prud\^encio}
\address{Coordination in Science and Technology - CCCT,\\
Federal University of Maranh\~ao - UFMA,\\ Campus Bacanga, 65080-805, S\~ao
Luis-MA, Brazil.}
\ead{prudencio.thiago@ufma.br}
\author{T. M. Rocha Filho}
\address{International Centre for Condensed Matter Physics - ICCMP \& Institute
of Physics,\\ University of Bras\'ilia - UnB, \\CP: 04455, 70919-970,
Bras\'ilia - DF, Brazil.}
\author{A. E. Santana}
\address{International Centre for Condensed Matter Physics - ICCMP \& Institute
of Physics, \\ University of Bras\'ilia - UnB, \\ CP: 04455, 70919-970,
Bras\'ilia - DF, Brazil.}

\begin{abstract}
Thermofield dynamics (TFD) approach is a real time quantum field method for dealing with finite temperature quantum states in a purified 
version of usual density operator formalism at finite temperature. In the domain of quantum information, 
TFD represents a quite promising direction for dealing with qubits under thermal influence and can also be associated to Gaussian states.
Here, we propose a generalized TFD mean expectation for the case of thermofield qubits considering the action of 
gate operators. We propose quantum teleportation protocols involving thermofield states, considering 
thermal-to-thermal and thermal-to-non-thermal transfering cases. In particular, we discuss the case in which 
Alice and Bob are at different temperatures. Action of gate operators on the result of the Mandel parameter for thermofields and 
on Gibbs-like density operators are also discussed. The no-cloning and non-broadcasting theorems in TFD are also considered and 
cases of superposed thermofield states and maps connecting thermofield vacua at different temperatures are also addressed and associated 
to metastable and non-equilibrium scenarios.
\end{abstract}

\begin{keyword}
Themofields, quantum states, qubits.
\end{keyword}

\end{frontmatter}


\section{Introduction}

Generation of thermal states by doubling the degrees of freedom in a Hilbert space accompanied by the action of a 
temperature dependent Bogoliubov transformation, thermofield dynamics (TFD) \cite{umezawabook,umezawabook2,dasbook}, is a natural approach for 
dealing with finite temperature quantum states in a purified version of usual density operator formalism in a finite temperature scenario. 
Proposed by Takahashi and Umezawa \cite{takahashi} as a real time quantum field theory at finite temperature, TFD has been applied in 
different contexts\cite{ojima,nakano1,nakahara,matsumoto2,matsumoto3,knight,suzuki,chu,kopf}, ranging high energy 
physics \cite{leblanc,landsman,kobes,silva,queiroz,balachandran,costa,leineker,belich}, quantum statistical mechanics \cite{tanabe,tay}, 
quantum optics \cite{barnett,mann,yamanaka,chaturvedi,ban,vourdas,matrasulov} 
and condensed matter\cite{matsumoto4,egorov,abreu}. Particularly, the thermofield state corresponding to the vacuum at finite
temperature has an expectation value equivalent to the equilibrium quantum
statistical measure. This state is associated to a thermal density operator at a given temperature, 
$T=\beta ^{-1}$. The physical correspondence of such thermofield vacuum with a mixed state associated to a given density operator 
was established previously and the physical meaning of the doubling has been fully identified (see \cite{santanaN} for a recent review). 
Considering an arbitrary mixed state, the non-tilde creation operators are identified with addition photon states, while tilde creation
operators are associated with subtraction photon states~\cite{ademirbook}.
On the other hand, realization of quantum information processing requires implementation of gate operations, incorporating 
transmission and state manipulation 
in a complete quantum computational scheme. One important role is represented by the quantum teleportation (QT), that dates back to 
Aharonov and Albert's result, in which nonlocality in a quantum system can be measured without violating causality \cite{aharonov}, the 
no-cloning theorem \cite{wooters} and the famous propose by Bennett et al.\cite{bennett} describing a protocol for
the transmission over spatial distances with reconstruction of a quantum
state, followed by an avalanche of other important theoretical proposes 
and experimental realizations (see for instance~\cite{vaidman,cirac,davidovich,braunstein1,bouwnmeester,furusawa,nielsen,
stenholm,zubairy,boschi,braunstein,reina,beenakker,fattal,ursin,biaozheng,cardoso,visser,beenakker,li,chen,sherson,zhang7,adesso}). 
In particular, the route for large scale quantum communication has been started with photonic and
ionic single qubits, polarized states of at 
least two-photons~\cite{zhang_goebel,manderbach}, increasing the range of fidelities above 0.8, covering distances above 
$\sim $ 140km~\cite{ma} and minimizing 
loss effects during transmission~\cite{strelchuk,bernien,nolleke,krauter,boixo}. A fundamental link to quantum computation was 
establised by Gottesman and Chuang proving that QT can be used as a universal primitive, reducing resource 
requirements for quantum computers and unifying known protocols for fault-tolerant quantum computation \cite{gottesman}. 

Bit-encondings using thermal logical bits, with proposes in thermal logical gates and thermal transistors \cite{wang2007}, reinforce 
the fact that the thermal properties can be used as a resource for transmission of information and computation. 

Quantum circuits incorporating incoherent resources \cite{santos}, fault-tolerant logical gates \cite{pastawski}, systems 
tolerant to decoherence arising from local noise \cite{brooks} and controllability of qubits\cite{zhang15} are examples of efforts 
to circunvent the inevitable presence of noise and environment effects in real quantum computation scenarios, justifying the search 
for new approaches for quantum information protocols. These routes can be extended in particular in the framework of quantum states 
at finite temperature, where TFD and other finite temperature methods can enter into play.

Indeed, TFD approach to quantum information represents a quite promising direction for
dealing with qubits under a thermal environment influence. In quantum information protocols, in particular QT, TFD has 
the possibility of dealing with non-locality and entanglement at finite temperature scenario making use of the algebraic
structure of such thermofield states. This route for the investigation of non-locality in thermal environments brings new features relating 
quantum information protocols in a thermofield scenario and have been explored in some recent proposes associated to 
maximally entangled states\cite{santana1,santana2,prudencio2}, no-cloning theorem \cite{prudencio1}, quantum gates \cite{trindade}. Thermofield
states are also used for description of eternal anti-de Sitter (AdS) black holes \cite{maldacena2,maldacena} with association to quantum 
complexity \cite{stanford} and can also be associated to 
Gaussian states \cite{weedbrook}.

In this work, the effect of temperature is implemented via TFD, by means of which we reconsider the expectation relations of thermal states for the 
case of thermofield qubits, deriving a generalized relation for this case under the action of gate operations. We also propose 
protocols where thermal-to-thermal and thermal-to-non-thermal quantum transfers and QT are realized with thermofield states. In particular, 
the case in which Alice and Bob are at different temperatures is also considered and discussed. 
We also discuss the TFD formulations involving the Mandel parameter and Gibbs like operators under the action of gate
operations. No-cloning theorem for thermofields \cite{prudencio1} and the problem of non-broadcasting 
\cite{barnum} in temperature dependent situations are investigated, where the connection among different thermofield states at
different temperatures and how such a method can be used in the domain of
metastable and non-equilibrium states are also discussed.

The paper is organized as follows: In section II, we discuss generalized thermofield mean expectations and the 
action of gate operations on a thermofield qubits. In section III, we propose QT of thermofield qubits. 
In section IV, we consider changing the Mandel parameter of a thermofield state under gate
operations. In Section V, we consider Gibbs-like density operators under gate operations. In section VI, we discuss the no-cloning theorem for TFD. 
In section VII, we consider maps connecting thermofield vacua, no-broadcasting theorem and 
superposition of thermofield vacua. We also consider
superpositions of thermofield  states at different temperatures and discuss their application in
 metastable and non-equilibrium scenarios. Finally, in
section VIII, we address our concluding remarks.

\section{Generalized thermofield mean expectations and the 
action of gate operations on a thermofield qubit}

We start with a superposition of thermofield states
\begin{eqnarray}
|\psi (\beta )\rangle =\sum_{j\in Z_{n+1}}a_{j}|j(\beta )\rangle,\label{1e}
\end{eqnarray}
where $a_{j}$ are arbitrary complex numbers satisfying $\sum_{j\in Z_{n+1}}|a_{j}|^{2}=1$ and the set 
$Z_{n+1}=\{0,1,...,n\}$ is the set of positive integer numbers mod $n+1$. In the case of $Z_{2}$, the set obeys simple algebra with
$1+1=0+0=0$, $1+0=0+1=1$. The states $|j(\beta )\rangle$ are $j$-order excitations from the thermofield vacuum 
$|0(\beta )\rangle$, by the action of a thermal creation operator \cite{ademirbook}. The association of the expectation value
of thermal vacuum with the statistical mean is given by the following relation 
\begin{eqnarray}
\langle \hat{\mathcal{O}}\rangle =\langle 0(\beta )|\hat{%
\mathcal{O}}|0(\beta )\rangle =Tr(\hat{\rho}^{th}\hat{\mathcal{O}}),
\end{eqnarray}
where $\hat{\mathcal{O}}$ is an operator acting on the non-tilde sector of the
thermofield and $\hat{\rho}^{th}$ is the associated thermal density operator. In the case of the bosonic state with associated number state $\hat{\mathcal{O}}=\hat{n}$, the thermofield vacuum implies
\begin{eqnarray}
\bar{n}=\langle \hat{n}\rangle =\langle 0(\beta )|\hat{a}^{\dagger}\hat{a}|0(\beta )\rangle =Tr(\hat{\rho}^{th}\hat{n}),
\end{eqnarray}
where $\hat{\rho}^{th}$ can be decomposed in terms of number states 
\begin{eqnarray}
\hat{\rho}^{th}=\sum_{n=0}^{\infty}\frac{\bar{n}^{n}}{(\bar{n}+1)^{n+1}}|n\rangle\langle n|.
\end{eqnarray}
For the case of associated modes with energy frequency $\omega$, the following Bose-Einstein distribution 
\begin{eqnarray}
\bar{n}=\frac{1}{1+e^{\beta\omega}}
\end{eqnarray}
is associated to the thermal state and maximizes the von Neumann entropy $S(\hat{\rho})=-Tr\left(\hat{\rho}\log\hat{\rho}\right)$,
\begin{eqnarray}
S(\hat{\rho}^{th})= \max S(\hat{\rho}).
\end{eqnarray}

Considering the state in Eq. (\ref{1e}), the expectation value for the observable $\hat{\mathcal{O}}$ can be defined by means of relation
\begin{eqnarray}
\langle \psi (\beta )|\hat{\mathcal{O}}|\psi (\beta )\rangle
=\sum_{j,j^{\prime }\in Z_{n}}a_{j}a_{j^{\prime }}^{\ast }\langle j(\beta )|%
\hat{\mathcal{O}}|j^{\prime }(\beta )\rangle , \label{2}
\end{eqnarray}
where $a_{j^{\prime }}^{\ast }$ is the complex conjugated of $a_{j^{\prime }}
$.

By restricting ourselves to the set $Z_{2}$, forming a thermofield qubit of the thermal vacuum and its first thermal excitation, the Bogoliubov relation $\hat{c}^{\dagger }=u(\beta )\hat{\mathbf{c}}%
^{\dagger }(\beta )+v(\beta )\widetilde{\mathbf{c}}(\beta )$, where $\hat{%
\mathbf{c}}(\beta )$ and $\widetilde{\mathbf{c}}(\beta )$ are the
corresponding non-tilde and tilde thermofield operators of annihilation, can be used to write the association between the excited non-tilde thermofield
and the thermofield vacuum by means of the following relation
\begin{eqnarray}
|1(\beta )\rangle =\frac{\hat{c}^{\dagger }}{u(\beta )}|0(\beta )\rangle . \label{tri8}
\end{eqnarray}
Using this expression, we rewrite Eq. (\ref{2}), $n=2$, explicitly in terms of
traces 

\begin{eqnarray}
\langle \hat{\mathcal{O}}\rangle _{\psi (\beta )} &=&\sum_{j,j^{\prime }\in
Z_{2}}a_{j}a_{j^{\prime }}^{\ast }\langle 0(\beta )|(\frac{\hat{c}}{u(\beta )%
})^{j}\hat{\mathcal{O}}(\frac{\hat{c}^{\dagger }}{u(\beta )})^{j^{\prime
}}|0(\beta )\rangle ,  \nonumber \\
&=&\sum_{j,j^{\prime }\in Z_{2}}\frac{a_{j}}{u(\beta )^{j}}\frac{%
a_{j^{\prime }}^{\ast }}{u(\beta )^{j^{\prime }}}\langle 0(\beta )|\hat{c}%
^{j}\hat{\mathcal{O}}\hat{c}^{\dagger j^{\prime }}|0(\beta )\rangle ,
\nonumber \\
&=&\sum_{j,j^{\prime }\in Z_{2}}\frac{a_{j}}{u(\beta )^{j}}\frac{%
a_{j^{\prime }}^{\ast }}{u(\beta )^{j^{\prime }}}Tr(\hat{\rho}\hat{c}^{j}%
\hat{\mathcal{O}}\hat{c}^{\dagger j^{\prime }}),  \nonumber \\
&=&\sum_{j,j^{\prime }\in Z_{2}}\frac{a_{j}}{u(\beta )^{j}}\frac{%
a_{j^{\prime }}^{\ast }}{u(\beta )^{j^{\prime }}}Tr(\hat{c}^{\dagger
j^{\prime }}\hat{\rho}\hat{c}^{j}\hat{\mathcal{O}}),
\end{eqnarray}%
where $\langle \hat{\mathcal{O}}\rangle _{\psi (\beta )}=\langle \psi (\beta
)|\hat{\mathcal{O}}|\psi (\beta )\rangle $. This mean value can be considered as taken in a mixture involving density matrices with particle addition.
Notice that this expectation can be rewritten in the form
\begin{equation}
\langle \hat{\mathcal{O}}\rangle _{\psi (\beta )}=Tr(\hat{\rho}_{\psi }\hat{%
\mathcal{O}})  \label{a3}
\end{equation}
where
\begin{equation}
\hat{\rho}_{\psi }=\sum_{j,j^{\prime }\in Z_{2}}\frac{a_{j}}{u(\beta )^{j}}%
\frac{a_{j^{\prime }}^{\ast }}{u(\beta )^{j^{\prime }}}\hat{c}^{\dagger
j^{\prime }}\hat{\rho}^{th}\hat{c}^{j}.  \label{a4}
\end{equation}%
This density operator is also written as
\begin{eqnarray}
\hat{\rho}_{\psi }=\frac{|a_{0}|^{2}}{u(\beta )^{2}}\hat{{c}}^{\dagger }\hat{%
\rho}^{th}\hat{{c}}+|a_{1}|^{2}\hat{\rho}^{th}+\frac{a_{0}^{\ast }a_{1}}{u(\beta )}%
\hat{\rho}^{th}\hat{{c}}+\frac{a_{0}^{\ast }a_{1}}{u(\beta )}\hat{{c}}^{\dagger }%
\hat{\rho}^{th}.
\end{eqnarray}
The expectation value given in Eq. (\ref{a3}) represents a generalization of
the thermofield expectation for the thermofield vacuum, corresponding to a
superposition of thermofields associated directly to a density operator as
given in Eq. (\ref{a4}).

The action of a non-thermalized gate operation on the state in Eq. (\ref{a4}) can be represented by an unitary operator $\hat{\mathcal{U}}_{G}$ defining
a new state
\[
\hat{\rho}_{\psi }^{(G)}=\hat{\mathcal{U}}_{G}\hat{\rho}_{\psi }\hat{%
\mathcal{U}}_{G}^{\dagger }.
\]%
Then, explicitly $\hat{\rho}_{\psi }^{(G)}$ reads
\begin{eqnarray}
\hat{\rho}_{\psi }^{(G)} &=&\frac{|a_{0}|^{2}}{u(\beta )^{2}}\hat{{\ c}}%
_{G}^{\dagger }\hat{\rho}_{G}\hat{{\ c}}_{G}+|a_{1}|^{2}\hat{\rho}_{G} +\frac{a_{0}^{\ast }a_{1}}{u(\beta )}\hat{\rho}_{G}\hat{{\ c}}_{G}+\frac{%
a_{0}^{\ast }a_{1}}{u(\beta )}\hat{{c}}_{G}^{\dagger }\hat{\rho}_{G},
\end{eqnarray}%
or, equivalently,
\begin{eqnarray}
\hat{\rho}_{\psi }^{(G)}=\sum_{j,j^{\prime }\in Z_{2}}\frac{a_{j}}{u(\beta
)^{j}}\frac{a_{j^{\prime }}^{\ast }}{u(\beta )^{j^{\prime }}}\hat{c}%
_{G}^{\dagger j^{\prime }}\hat{\rho}_{G}\hat{c}_{G}^{j}. \label{kj}
\end{eqnarray}
where now
\begin{eqnarray}
\hat{\rho}_{G} &=&\hat{\mathcal{U}}_{G}\hat{\rho}\hat{\mathcal{U}}%
_{G}^{\dagger }, \\
\hat{c}_{G} &=&\hat{\mathcal{U}}_{G}\hat{c}\hat{\mathcal{U}}_{G}^{\dagger }.
\label{gat1}
\end{eqnarray}%
This result implies that the action of a gate operation in the density
operator associated in the thermofield qubit is obtained by the action of
the same gate operation simultaneously in the density operator associated to
the thermofield vacuum and the non-thermal creation and annihilation
operators.

Taking the trace associated to the mean expectation of the operator $\hat{%
\mathcal{O}}$, we have, considering the unitarity of $\hat{\mathcal{U}}_{G}$,
\begin{eqnarray*}
Tr(\hat{\rho}_{\psi }^{(G)}\hat{\mathcal{O}}) &=&\frac{|a_{0}|^{2}}{u(\beta )%
}\langle c\hat{\mathcal{U}}_{G}^{\dagger }\mathcal{O}\hat{\mathcal{U}}%
_{G}c^{\dagger }\rangle  \\
&&+|a_{1}|^{2}\langle \hat{\mathcal{U}}_{G}^{\dagger }\mathcal{O}\hat{%
\mathcal{U}}_{G}\rangle +\frac{a_{0}^{\ast }a_{1}}{u(\beta )}\langle \hat{{\
c}}\hat{\mathcal{U}}_{G}^{\dagger }\mathcal{O}\hat{\mathcal{U}}_{G}\rangle
\\
&&+\frac{a_{0}^{\ast }a_{1}}{u(\beta )}\langle \hat{\mathcal{U}}%
_{G}^{\dagger }\mathcal{O}\hat{\mathcal{U}}_{G}\hat{{\ c}}^{\dagger }\rangle
.
\end{eqnarray*}%
Now let us consider the thermofield state resulting from the density
operator under gate operation, by means of the identification
\begin{eqnarray}
\langle \psi ^{(G)}(\beta )|\hat{\mathcal{O}}|\psi ^{(G)}(\beta )\rangle =Tr(%
\hat{\rho}_{\psi }^{(G)}\hat{\mathcal{O}}), \label{ga3}
\end{eqnarray}
where $|\psi ^{(G)}(\beta )\rangle $ is the modified thermofield state
resulting from the gate operation on the thermofield qubit. We can check the
following
\[
|\psi ^{(G)}(\beta )\rangle =\sum_{j\in Z_{2}}a_{j}|j_{G}(\beta )\rangle ,
\]%
where
\begin{eqnarray}
|1_{G}(\beta )\rangle =\frac{\hat{c}_{G}^{\dagger }}{u(\beta )}|0_{G}(\beta
)\rangle \label{1g}
\end{eqnarray}
and
\begin{eqnarray}
\langle 0_{G}(\beta )|\hat{\mathcal{O}}|0_{G}(\beta )\rangle =Tr(\rho _{G}%
\hat{\mathcal{O}}).
\end{eqnarray}
Finally, we can implement the correspondence with thermofield vacuum by
means of
\[
Tr(\rho _{G}\hat{\mathcal{O}})=Tr(\rho \hat{\mathcal{U}}_{G}^{\dagger }\hat{%
\mathcal{O}}\hat{\mathcal{U}}_{G})=\langle 0(\beta )|\hat{\mathcal{U}}%
_{G}^{\dagger }\hat{\mathcal{O}}\hat{\mathcal{U}}_{G}|0(\beta )\rangle .
\]%
From this, we can finally make the identification
\begin{eqnarray}
|0_{G}(\beta )\rangle =\hat{\mathcal{U}}_{G}|0(\beta )\rangle .\label{0g}
\end{eqnarray}
This result implies that the thermofield vacuum can be directly operated
by the gate operation and that Eqs. (\ref{1g}) and (\ref{0g}) determine
fully the action of the gate operator on the thermofield qubit, in
complete agreement with the rearrangement of density operator and
annihilation operators under the gate.

There are some subtleties that deserves to be analyzed. Under the gate
operation the Bogoliubov relation is written as
\begin{eqnarray}
\hat{c}_{G}^{\dagger }=u(\beta )\hat{\mathbf{c}}_{G}^{\dagger }(\beta
)+v(\beta )\widetilde{\mathbf{c}}_{G}(\beta ),
\end{eqnarray}
with
\begin{eqnarray}
\hat{\mathbf{c}}_{G}(\beta ) &=&\hat{\mathcal{U}}_{G}\hat{\mathbf{c}}(\beta )%
\hat{\mathcal{U}}_{G}^{\dagger }, \\
\widetilde{\mathbf{c}}_{G}(\beta ) &=&\hat{\mathcal{U}}_{G}\widetilde{%
\mathbf{c}}(\beta )\hat{\mathcal{U}}_{G}^{\dagger }.
\end{eqnarray}%
The action into the thermofield vacuum now depends on the result of the
action of the gate operation
\begin{eqnarray}
\hat{c}_{G}^{\dagger }|0(\beta )\rangle =u(\beta )\hat{\mathbf{c}}%
_{G}^{\dagger }(\beta )|0(\beta )\rangle +v(\beta )\widetilde{\mathbf{c}}%
_{G}(\beta )|0(\beta )\rangle . \label{u12w}
\end{eqnarray}
We also have in the gate operated state , Eqs. (\ref{0g}) and (\ref{1g}),
\begin{eqnarray}
\hat{c}_{G}^{\dagger }|0_{G}(\beta )\rangle  &=&u(\beta )\hat{\mathbf{c}}%
_{G}^{\dagger }(\beta )|0_{G}(\beta )\rangle +v(\beta )\widetilde{\mathbf{c}}%
_{G}(\beta )|0_{G}(\beta )\rangle .  \nonumber \\
&=&u(\beta )|1_{G}(\beta )\rangle .  \label{yi}
\end{eqnarray}%
Then the state in Eq. (\ref{1g}) reads as
\begin{eqnarray}
|1_{G}(\beta )\rangle  &=&\hat{\mathbf{c}}_{G}^{\dagger }(\beta
)|0_{G}(\beta )\rangle +\frac{v(\beta )}{u(\beta )}\widetilde{\mathbf{c}}%
_{G}(\beta )|0_{G}(\beta )\rangle .  \nonumber \\
&&
\end{eqnarray}%
From Eqs. (\ref{u12w}) and (\ref{yi}),   the following commutation relations
are fulfilled
\[
\lbrack \hat{\mathcal{U}}_{G},\hat{c}_{G}^{\dagger }]=u(\beta )[\hat{%
\mathcal{U}}_{G},\hat{\mathbf{c}}^{\dagger }(\beta )]+v(\beta )[\hat{%
\mathcal{U}}_{G},\widetilde{\mathbf{c}}(\beta )].
\]%
Finally,   from Eqs. (\ref{1g}) and (\ref{tri8}), one has
\begin{eqnarray}
|1_{G}(\beta )\rangle  &=&\frac{\hat{c}_{G}^{\dagger }}{u(\beta )}\hat{%
\mathcal{U}}_{G}|0(\beta )\rangle   \nonumber \\
&=&\hat{\mathcal{U}}_{G}\frac{\hat{c}^{\dagger }}{u(\beta )}|0(\beta
)\rangle   \nonumber \\
&=&\hat{\mathcal{U}}_{G}|1(\beta )\rangle .
\end{eqnarray}%
This result shows that the action of the gate operator in the first excited
thermofield state coincides with the first excited gate operated state.

\section{Teleportation of thermofield qubits}

Let us consider as a qubit state the following superposition of thermofield
states, i. e., a thermofield qubit,
\begin{eqnarray}
|\psi (\beta )\rangle _{A}=\sum_{j\in Z_{2}}a_{j}|j(\beta )\rangle _{A},\label{1}
\end{eqnarray}
where $a_{j}$ are unknown complex numbers. The states $|0(\beta )\rangle $ and $|1(\beta
)\rangle $ are thermofield vacuum and its first excitation, respectively
\cite{ademirbook}. The state given in Eq. (\ref{1}) is prepared in
Alice's lab.

A quantum channel composed of two entangled thermofield states, with a
particle $C$ belonging to Alice and a particle $B$ belonging to Bob can
be described by the following superposition of thermofield states
\[
|\psi (\beta )\rangle _{BC}=\sum_{j\in Z_{2}}(-1)^{j}|j(\beta )\rangle
_{B}|(j+1)(\beta )\rangle _{C}.
\]%
The whole state can be written as

\begin{eqnarray*}
|\psi (\beta )\rangle _{ABC} &=&\sum_{j,j^{\prime }\in
Z_{2}}(-1)^{j}a_{j^{\prime }}|j^{\prime }(\beta )\rangle _{A}|j(\beta )\rangle _{B}|j+1(\beta )\rangle _{C}.
\end{eqnarray*}

The Alice's system composed of particles $A$ and $C$ can be conveniently
rewritten in terms of a Bell basis of thermofield states, defined by
\begin{eqnarray}
|\Psi _{s}(\beta )\rangle _{AC} &=&\sum_{j^{\prime \prime }\in
Z_{2}}s^{j^{\prime \prime }}|j^{\prime \prime }(\beta )\rangle
_{A}|(j^{\prime \prime }+1)(\beta )\rangle _{C}, \\
|\Phi _{s}(\beta )\rangle _{AC} &=&\sum_{j^{\prime \prime }\in
Z_{2}}s^{j^{\prime \prime }}|j^{\prime \prime }(\beta )\rangle
_{A}|j^{\prime \prime }(\beta )\rangle _{C},
\end{eqnarray}%
where $s=\pm 1$. With this in view, Alice projects the $AC$ state in one
of these Bell states, achieving at one of the following states
\begin{eqnarray}
_{AC}\langle \Psi _{s}(\beta )|\psi _{s^{\prime }}(\beta )\rangle
_{ABC}=\sum_{j\in Z_{2}}(-1)^{j}a_{j}s^{j}|j(\beta )\rangle _{B},
\label{heu1}
\end{eqnarray}
\begin{eqnarray}
_{AC}\langle \Phi _{s}(\beta )|\psi _{s^{\prime }}(\beta )\rangle
_{ABC}=\sum_{j\in Z_{2}}(-1)^{j}a_{j+1}s^{j+1}|j(\beta )\rangle _{B},
\label{heu2}
\end{eqnarray}
where the orthonormality relations lead to $j^{\prime \prime }=j^{\prime }=j$
in  Eq. (\ref{heu1}) and $j^{\prime \prime }=j^{\prime }=j+1$ in Eq. (\ref{heu2}).

Once Alice's decide in which basis realizing her measurement, she tells
to Bob using a classical device what was her procedure. With such
information Bob makes a corresponding projection choosing one of the
following projectors
\begin{eqnarray}
P_{s,s^{\prime}}^{\Psi}&=&\sum_{j\in
Z_{2}}(-1)^{j}s^{j}|j(\beta)\rangle\langle j(\beta)|,  \label{be3} \\
P_{s,s^{\prime}}^{\Phi}&=&\sum_{j\in
Z_{2}}(-1)^{j}s^{j+1}|j+1(\beta)\rangle\langle j(\beta)|.  \label{be4}
\end{eqnarray}
By implementing this procedure, Bob achieve at the following state
\begin{eqnarray}
|\psi(\beta)\rangle_{B} = \sum_{j\in Z_{2}} a_{j}|j(\beta)\rangle_{B},
\end{eqnarray}
corresponding to the teleportation of the thermofield state (\ref{1}).

This procedure is a fidelity $1$ quantum teleportation and makes use only
of the algebraic properties of the thermofield states \cite{ademirbook}.
Due to this structure, Alice and Bob can also be in different
temperatures or thermal baths, $\beta$ and $\beta'$, such that the final
state of Bob is a thermofield qubit of temperature ${\beta'}^{-1}$ while
Alice's has temperature $\beta^{-1}$. We will consider this point below.

Let us now consider the following initial states
\begin{eqnarray}
|\psi \rangle _{00A} &=&\sum_{j\in Z_{2}}a_{j}|j,\tilde{0}\rangle _{A}, \nonumber \\
|\psi \rangle _{11A} &=&\sum_{j\in Z_{2}}a_{j}|j,\tilde{1}\rangle _{A}, \nonumber\\
|\psi \rangle _{01A} &=&\sum_{j\in Z_{2}}a_{j}|j,\widetilde{j+1}\rangle _{A},\nonumber
\\
|\psi \rangle _{10A} &=&\sum_{j\in Z_{2}}a_{j}|j,\tilde{j}\rangle _{A}, \nonumber
\end{eqnarray}%
defined in the space $\mathcal{H}\otimes \tilde{\mathcal{H}}$, but not in
contact with a heat bath. 
Notice that although the states with subscripts $00$, $11$ are separable,
the states $01$ and $10$ are entangled in  $\mathcal{H}\otimes \tilde{%
\mathcal{H}}$. We can simplify in a matrix form
\begin{eqnarray}
\left ( \begin{array}{cc}
     {|\psi\rangle_{00A}} & {|\psi\rangle_{01A}} \\
     {|\psi\rangle_{10A}} & {|\psi\rangle_{11A}} \\
   \end{array} \right ).
\end{eqnarray}
Alice has two particles that are not in a thermal bath, i.e., states at zero
temperature. The Alice's particle $C$ is shared in a quantum channel with
Bob, that has a thermofield state, i.e., Bob's particle is in contact to a
heat bath. Let us consider the quantum channel given by the following state
\[
|\psi (\beta )\rangle _{BC}=\sum_{j^{\prime }\in Z_{2}}(-1)^{j^{\prime
}}|j^{\prime }(\beta )\rangle _{B}|j^{\prime }+1,\widetilde{j^{\prime }+1}%
\rangle _{C}.
\]%
The total state for each one of the initial states are given by

\begin{eqnarray}
|\psi_{00}\rangle_{ABC}&=& \sum_{j,j^{\prime}\in
Z_{2}}(-1)^{j^{\prime}}a_{j}|j,\tilde{0}\rangle_{A}|j^{\prime}(\beta)%
\rangle_{B}|j^{\prime}+1,\widetilde{j^{\prime}+1}\rangle_{C},  \nonumber \\
|\psi_{11}\rangle_{ABC}&=& \sum_{j,j^{\prime}\in
Z_{2}}(-1)^{j^{\prime}}a_{j}|j,\tilde{1}\rangle_{A}|j^{\prime}(\beta)%
\rangle_{B}|j^{\prime}+1,\widetilde{j^{\prime}+1}\rangle_{C}, \nonumber \\
|\psi_{01}\rangle_{ABC}&=& \sum_{j,j^{\prime}\in
Z_{2}}(-1)^{j^{\prime}}a_{j}|j,\widetilde{j+1}\rangle_{A}|j^{\prime}(\beta)%
\rangle_{B}|j^{\prime}+1,\widetilde{j^{\prime}+1}\rangle_{C}, \nonumber\\
|\psi_{10}\rangle_{ABC}&=& \sum_{j,j^{\prime}\in
Z_{2}}(-1)^{j^{\prime}}a_{j}|j,\tilde{j}\rangle_{A}|j^{\prime}(\beta)%
\rangle_{B}|j^{\prime}+1,\widetilde{j^{\prime}+1}\rangle_{C}.\nonumber
\end{eqnarray}
We have convenient $00$-Bell basis for each one of the above $AC$-subsystems
\begin{eqnarray}
|b_{s,00}^{(1)}\rangle _{AC} &=&\sum_{j^{\prime \prime }\in
Z_{2}}s^{j^{\prime \prime }}|j^{\prime \prime },\tilde{0}\rangle
_{A}|j^{\prime \prime }+1,\widetilde{j^{\prime \prime }+1}\rangle _{C}, \\
|b_{s,00}^{(2)}\rangle _{AC} &=&\sum_{j^{\prime \prime }\in
Z_{2}}s^{j^{\prime \prime }}|j^{\prime \prime },\tilde{0}\rangle
_{A}|j^{\prime \prime },\widetilde{j^{\prime \prime }}\rangle _{C},
\end{eqnarray}%
where $s=\pm $.

Alice's projections in each one of this states will give
\begin{eqnarray}
\langle b_{s,00}^{(1)}|\psi _{00}\rangle _{ABC} &=&\sum_{j\in
Z_{2}}(-1)^{j}s^{j}a_{j}|j(\beta )\rangle _{B}, \\
\langle b_{s,00}^{(2)}|\psi _{00}\rangle _{ABC} &=&\sum_{j\in
Z_{2}}(-1)^{j}s^{j+1}a_{j+1}|j(\beta )\rangle _{B}.
\end{eqnarray}%
The $11$-Bell basis is given by
\begin{eqnarray}
|b_{s,11}^{(1)}\rangle _{AC} &=&\sum_{j^{\prime \prime }\in
Z_{2}}s^{j^{\prime \prime }}|j^{\prime \prime },\tilde{1}\rangle
_{A}|j^{\prime \prime }+1,\widetilde{j^{\prime \prime }+1}\rangle _{C}, \\
|b_{s,11}^{(2)}\rangle _{AC} &=&\sum_{j^{\prime \prime }\in
Z_{2}}s^{j^{\prime \prime }}|j^{\prime \prime },\tilde{1}\rangle
_{A}|j^{\prime \prime },\widetilde{j^{\prime \prime }}\rangle _{C}.
\end{eqnarray}%
Such that the projections lead to
\begin{eqnarray}
\langle b_{s,11}^{(1)}|\psi _{11}\rangle _{ABC} &=&\sum_{j\in
Z_{2}}(-1)^{j}s^{j}a_{j}|j(\beta )\rangle _{B}, \\
\langle b_{s,11}^{(2)}|\psi _{11}\rangle _{ABC} &=&\sum_{j\in
Z_{2}}(-1)^{j}s^{j+1}a_{j+1}|j(\beta )\rangle _{B}.
\end{eqnarray}%
The $01$-Bell basis is given by
\begin{eqnarray}
|b_{s,01}^{(1)}\rangle _{AC} &=&\sum_{j^{\prime \prime }\in
Z_{2}}s^{j^{\prime \prime }}|j^{\prime \prime },\tilde{j^{\prime \prime }+1}%
\rangle _{A}|j^{\prime \prime }+1,\widetilde{j^{\prime \prime }+1}\rangle
_{C},  \nonumber \\
&& \\
|b_{s,01}^{(2)}\rangle _{AC} &=&\sum_{j^{\prime \prime }\in
Z_{2}}s^{j^{\prime \prime }}|j^{\prime \prime },\tilde{j^{\prime \prime }+1}%
\rangle _{A}|j^{\prime \prime },\widetilde{j^{\prime \prime }}\rangle _{C}.
\end{eqnarray}%
The projections lead to
\begin{eqnarray}
\langle b_{s,01}^{(1)}|\psi _{01}\rangle _{ABC} &=&\sum_{j\in
Z_{2}}(-1)^{j}s^{j}a_{j}|j(\beta )\rangle _{B}, \\
\langle b_{s,01}^{(2)}|\psi _{01}\rangle _{ABC} &=&\sum_{j\in
Z_{2}}(-1)^{j}s^{j+1}a_{j+1}|j(\beta )\rangle _{B}.
\end{eqnarray}%
Finally, the $10$-Bell basis is given by
\begin{eqnarray}
|b_{s,10}^{(1)}\rangle _{AC} &=&\sum_{j^{\prime \prime }\in
Z_{2}}s^{j^{\prime \prime }}|j^{\prime \prime },\tilde{j^{\prime \prime }}%
\rangle _{A}|j^{\prime \prime }+1,\widetilde{j^{\prime \prime }+1}\rangle
_{C},  \nonumber \\
&& \\
|b_{s,10}^{(2)}\rangle _{AC} &=&\sum_{j^{\prime \prime }\in
Z_{2}}s^{j^{\prime \prime }}|j^{\prime \prime },\tilde{j^{\prime \prime }}%
\rangle _{A}|j^{\prime \prime },\widetilde{j^{\prime \prime }}\rangle _{C},
\end{eqnarray}%
with the projections leading to
\begin{eqnarray}
\langle b_{s,10}^{(1)}|\psi _{10}\rangle _{ABC} &=&\sum_{j\in
Z_{2}}(-1)^{j}s^{j}a_{j}|j(\beta )\rangle _{B}, \\
\langle b_{s,10}^{(2)}|\psi _{10}\rangle _{ABC} &=&\sum_{j\in
Z_{2}}(-1)^{j}s^{j+1}a_{j+1}|j(\beta )\rangle _{B}.
\end{eqnarray}%
When Alice tells to Bob in which basis she realized the projection, Bob
can apply one of the projectors given in Eq. (\ref{be3}) or Eq.
(\ref{be4}), achieving in the thermofield qubit
\[
|\psi (\beta )\rangle _{B}=\sum_{j\in Z_{2}}a_{j}|j(\beta )\rangle _{B}.
\]%
The same idea can be used when Alice and Bob are in two different thermal
baths with a shared quantum channel with sub-states at different
temperatures. The presence of non-locality associated to the entanglement
of quantum states at different temperatures is a still not well explored
subject. Since the thermal interaction between two subsystems is a local
effect, the non-local channel at different temperatures can be used for
realizing quantum information protocols. In this point, the thermofield
states give a clear difference between the non-local effect, given by the
entanglement of the states at different temperatures, and the local
effect that is the temperature associated to each subsystem.
\newline
\newline

\section{Changing the Mandel parameter of thermofield state under Gate
operation}

Let us consider the Mandel parameter in the case of thermofields, as
discussed in a recent work \cite{trindade}. Here we will consider how this
parameter can be changed under gate operation, in particular in the action
in the thermofield vacuum and in a thermofield qubit. This quantity is
described in terms of traces by means of the following
\[
\mathcal{Q}=\frac{Tr(\hat{\rho}(\hat{n}^{2}-\hat{n}))-[Tr(\hat{\rho}\hat{n}%
)]^{2}}{Tr(\hat{\rho}\hat{n})},
\]%
where $\hat{n}=\hat{c}^{\dagger }\hat{c}$.

Under a gate operation, we have
\[
\mathcal{Q}_{G}=\frac{Tr(\hat{\rho}_{G}(\hat{n}^{2}-\hat{n}))-[Tr(\hat{\rho}%
_{G}\hat{n})]^{2}}{Tr(\hat{\rho}_{G}\hat{n})},
\]%
Using Eq. (\ref{ga3}), then
\begin{eqnarray}
\mathcal{Q}_{G} &=&\frac{\langle 0_{G}(\beta )|(\hat{n}^{2}-\hat{n}%
)|0_{G}(\beta )\rangle -\langle 0_{G}(\beta )|\hat{n}|0_{G}(\beta )\rangle
^{2}}{\langle 0_{G}(\beta )|\hat{n}|0_{G}(\beta )\rangle }.  \nonumber \\
&&
\end{eqnarray}%
From Eq. (\ref{0g}) we can also write
\begin{eqnarray}
\langle 0_{G}(\beta )|\hat{n}|0_{G}(\beta )\rangle  &=&\langle 0(\beta )|%
\hat{\mathcal{U}}_{G}^{\dagger }\hat{n}\hat{\mathcal{U}}_{G}|0(\beta
)\rangle   \nonumber \\
&=&\langle \hat{\mathcal{U}}_{G}^{\dagger }\hat{n}\hat{\mathcal{U}}%
_{G}\rangle .
\end{eqnarray}%
The action of the gate operation on the thermofield vacuum produces a change
in the Mandel parameter given, finally, by
\[
\mathcal{Q}_{G}=\frac{\langle \hat{\mathcal{U}}_{G}^{\dagger }(\hat{n}^{2}-%
\hat{n})\hat{\mathcal{U}}_{G}\rangle -\langle \hat{\mathcal{U}}_{G}^{\dagger
}\hat{n}\hat{\mathcal{U}}_{G}\rangle ^{2}}{\langle \hat{\mathcal{U}}%
_{G}^{\dagger }\hat{n}\hat{\mathcal{U}}_{G}\rangle }.
\]%
This result shows that the modified Mandel parameter depends on the gate
operation $\hat{\mathcal{U}}_{G}$.

Now let us analyse the situation with the thermofield qubit. We can write
the corresponding Mandel Parameter in the following way
\begin{eqnarray}
\mathcal{Q}_{G}^{\psi } &=&\frac{\langle \psi ^{(G)}(\beta )|(\hat{n}^{2}-%
\hat{n})|\psi ^{(G)}(\beta )\rangle }{\langle \psi ^{(G)}(\beta )|\hat{n}%
|\psi ^{(G)}(\beta )\rangle }  \nonumber \\
&&-\frac{\langle \psi ^{(G)}(\beta )|\hat{n}|\psi ^{(G)}(\beta )\rangle ^{2}%
}{\langle \psi ^{(G)}(\beta )|\hat{n}|\psi ^{(G)}(\beta )\rangle },
\end{eqnarray}%
which, by using  Eq. (\ref{ga3}), leads to
\[
\mathcal{Q}_{G}^{\psi }=\frac{Tr(\hat{\rho}_{\psi }^{(G)}(\hat{n}^{2}-\hat{n}%
))-Tr(\hat{\rho}_{\psi }^{(G)}\hat{n})^{2}}{Tr(\hat{\rho}_{\psi }^{(G)}\hat{n%
})},
\]%
Substituting the state given in Eq. (\ref{kj}), we  write

\begin{eqnarray*}
\mathcal{Q}_{G}^{\psi } &=&\left\{ \sum_{j,j^{\prime }\in Z_{2}}\frac{a_{j}}{%
u(\beta )^{j}}\frac{a_{j^{\prime }}^{\ast }}{u(\beta )^{j^{\prime }}}\langle
\hat{\mathcal{U}}_{G}^{\dagger }\hat{c}_{G}^{j}(\hat{n}^{2}-\hat{n})\hat{c}%
_{G}^{\dagger j^{\prime }}\hat{\mathcal{U}}_{G}\rangle \right.  \\
&&\left. -\left[ \sum_{j,j^{\prime }\in Z_{2}}\frac{a_{j}}{u(\beta )^{j}}%
\frac{a_{j^{\prime }}^{\ast }}{u(\beta )^{j^{\prime }}}\langle \hat{\mathcal{%
U}}_{G}^{\dagger }\hat{c}_{G}^{j}\hat{n}\hat{c}_{G}^{\dagger j^{\prime }}%
\hat{\mathcal{U}}_{G}\rangle \right] ^{2}\right\}  \\
&&\times \frac{1}{\sum_{j,j^{\prime }\in Z_{2}}\frac{a_{j}}{u(\beta )^{j}}%
\frac{a_{j^{\prime }}^{\ast }}{u(\beta )^{j^{\prime }}}\langle \hat{\mathcal{%
U}}_{G}^{\dagger }\hat{c}_{G}^{j}\hat{n}\hat{c}_{G}^{\dagger j^{\prime }}%
\hat{\mathcal{U}}_{G}\rangle },
\end{eqnarray*}%
This result shows that the action of the gate operation in a thermofield
qubit produces a change in the Mandel parameter corresponding to the action
of the unitary operator $\hat{\mathcal{U}}_{G}$ in the density operator
associated to the thermofield qubit, producing changes in the expectation
values. 

\section{Gibbs-like density operator and gate operator effect}

As it is has been shown earlier \cite{santana3}, a Hamiltonian in
the simple form $\hat{H}=\omega \hat{n}$ has in its thermofield form a
thermofield vacuum associated with a corresponding Gibbs-like density
operator $\hat{\rho}=e^{-\beta \omega \hat{n}}/\mathcal{Z}$, with $\mathcal{Z%
}=Tr(e^{-\beta \hat{H}})$. Using the results above we have that the modified
state under the action of a gate operation is given by
\[
\hat{\rho}_{G}=\frac{1}{\mathcal{Z}}\hat{\mathcal{U}}_{G}\exp {(-\beta
\omega \hat{n})}\hat{\mathcal{U}}_{G}^{\dagger }.
\]%
The corresponding thermofield qubit is now written in following way
\begin{eqnarray}
\hat{\rho}_{\psi }^{(G)} &=&\sum_{j,j^{\prime }\in Z_{2}}\frac{a_{j}}{%
u(\beta )^{j}}\frac{a_{j^{\prime }}^{\ast }}{u(\beta )^{j^{\prime }}}\hat{c}%
_{G}^{\dagger j^{\prime }}\frac{1}{\mathcal{Z}}\hat{\mathcal{U}}_{G}\exp {%
(-\beta \omega \hat{n})}\hat{\mathcal{U}}_{G}^{\dagger }\hat{c}_{G}^{j}.
\nonumber \\
&&
\end{eqnarray}%
Consider this problem in the context of a spin $1/2$ system. In terms of the
basis of spin, we can write \cite{ademirbook}
\[
\hat{\rho}=\frac{1}{\mathcal{Z}}e^{-\beta \omega \hat{S}_{0}}|\frac{1}{2}%
\rangle \langle \frac{1}{2}|+\frac{1}{\mathcal{Z}}e^{-\beta \omega \hat{S}%
_{0}}|-\frac{1}{2}\rangle \langle -\frac{1}{2}|,
\]%
such that $\hat{S}_{0}|\sigma \rangle =\sigma |\sigma \rangle $, $\sigma
=\pm 1/2$. Taking a gate operation as a Hadamard operation over the states,
we have $\hat{\mathcal{U}}_{Hadamard}|\pm \frac{1}{2}\rangle =\frac{1}{\sqrt{%
2}}\left( |\frac{1}{2}\rangle \pm |-\frac{1}{2}\rangle \right) $ leading to
the following modified state under this gate operation
\begin{eqnarray*}
\hat{\rho}^{(Hadamard)} &=&\frac{1}{2\mathcal{Z}}\sum_{s=\pm }\exp {(\frac{%
s\beta \omega }{2})}\times \left( |\frac{1}{2}\rangle +s|-\frac{1}{2}\rangle \right) \left( \langle\frac{1}{2}| +s\langle-\frac{1}{2}|\right) .
\end{eqnarray*}%
It is also important to notice that this operation is reversible, such
that we can restore the original state by applying a second Hadammard
gate
\[
\hat{\rho}=\hat{\mathcal{U}}_{Hadamard}\hat{\rho}^{(Hadamard)}\hat{\mathcal{U%
}}_{Hadamard}^{\dagger }.
\]%
This implies that one can also start with an operated gate state and recover
the corresponding thermofield state by applying an adequate reversible gate.

\section{No-cloning for thermofields}

Consider the action of the tilde conjugation defined by the action of a
doubling map
\begin{eqnarray}
\mathcal{D}_{TFD}(|0\rangle )=|0,\tilde{0}\rangle. \label{31}
\end{eqnarray}
There is an association between the doubling procedure in TFD and the
no-cloning theorem \cite{prudencio1}. The doubling procedure has the same
characteristics involved in the no-cloning theorem: it cannot be realized
unitarily for an arbitrary superposition state, since the requirement of
linearity cannot be achieved. As such, the extension
\begin{eqnarray}
\mathcal{D}_{TFD}(|\psi \rangle )=|\psi ,\tilde{\psi}\rangle, \label{12}
\end{eqnarray}
where $|\psi \rangle $ is a qubit state, is not implemented via unitary
operation. This does not constitute a problem for TFD itself since it starts
from the vacuum and generates the doubling vacuum for applying a temperature
dependent Bogoliubov operation. However, an extension of the method starting
from the doubling of arbitrary states is forbidden under linearity
requirement. Consequently,  Eq.  (\ref{12}) is not valid in general, but Eq.
(\ref{31}) is fully consistent.

Another important point here is the no-cloning theorem for thermofield
qubits. Consider a cloning map
\begin{eqnarray}
\mathcal{C}(\ast )=\ast \otimes \ast , \label{c1}
\end{eqnarray}
where $\ast $ is an arbitrary algebraic quantity. Let us consider the
thermofield qubit {\it inside} this map. We verify that
\begin{eqnarray}
\mathcal{C}(\sum_{j\in Z_{2}}a_{j}|j(\beta )\rangle ) &=&(\sum_{j\in
Z_{2}}a_{j}|j(\beta )\rangle )\otimes (\sum_{j^{\prime }\in
Z_{2}}a_{j^{\prime }}|j^{\prime }(\beta )\rangle ).  \nonumber \\
&=&\sum_{j,j^{\prime }\in Z_{2}}a_{j}a_{j^{\prime }}|j(\beta )\rangle
\otimes |j^{\prime }(\beta )\rangle .  \label{nc1}
\end{eqnarray}%
But this map is not linear since the linearity requires
\[
\sum_{j\in Z_{2}}a_{j}\mathcal{C}(|j(\beta )\rangle )=\sum_{j\in
Z_{2}}a_{j}|j(\beta )\rangle \otimes |j(\beta )\rangle ,
\]%
that is not equivalent to Eq. (\ref{nc1}):
\[
\mathcal{C}(\sum_{j\in Z_{2}}a_{j}|j(\beta )\rangle )\neq \sum_{j\in
Z_{2}}a_{j}\mathcal{C}(|j(\beta )\rangle ).
\]%
This result implies that a thermofield qubit cannot be cloned, in complete
agreement with the no-cloning theorem, at zero temperature.

\section{Maps for connecting thermofield vacua, no-broadcasting theorem and 
superposition of thermofield vacua}

Consider the following map
\begin{eqnarray}
\mathcal{T}(\hat{\rho})= \hat{\rho}^{\prime},  \label{tpo}
\end{eqnarray}
such that
\begin{eqnarray}
\langle 0(\beta)|\hat{\mathcal{O}}| 0(\beta)\rangle = Tr(\hat{\rho}\hat{%
\mathcal{O}}),
\end{eqnarray}
and
\begin{eqnarray}
\langle 0(\beta^{\prime})|\hat{\mathcal{O}}| 0(\beta^{\prime})\rangle = Tr(%
\hat{\rho}^{\prime}\hat{\mathcal{O}}).
\end{eqnarray}
This map take a thermofield vacuum associated to a given temperature ${\beta}%
^{-1}$ and leads to another thermofield vacuum associated to other
temperature ${\beta^{\prime}}^{-1}$.

In particular, the zero temperature state leads to
\begin{eqnarray}
\langle 0(\infty)|\hat{\mathcal{O}}| 0(\infty)\rangle = Tr(|0\rangle\langle
0|\hat{\mathcal{O}}).
\end{eqnarray}
Physically, this process has to be associated to an exchange of thermal
baths.

A whole density matrix incorporating all the associated temperatures can be
described by
\[
\hat{\rho}^{(\mathcal{T})}=\int d\beta \mu _{\beta }\hat{\rho}_{\beta }.
\]%
such that
\[
Tr(\hat{\rho}^{(\mathcal{T})}\hat{\mathcal{O}})=\int d\beta \mu _{\beta
}\langle 0(\beta )|\hat{\mathcal{O}}|0(\beta )\rangle .
\]%
This state discribes the passage for all temperatures, and the association to
a given temperature $\beta _{0}$ comes from the association $\mu _{\beta
}=\delta (\beta -\beta _{0})$ and
\[
\hat{\rho}_{\beta _{0}}=\int d\beta \delta (\beta -\beta _{0})\hat{\rho}%
_{\beta }.
\]%
We now return to the map in Eq. (\ref{tpo}). Applying in the above state it
leads to
\[
\mathcal{T}(\hat{\rho}_{\beta _{0}})=\int d\beta \tilde{\mathcal{T}}(\delta
(\beta -\beta _{0}))\hat{\rho}_{\beta }=\hat{\rho}_{\beta ^{\prime }},
\]%
and consequently,
\[
\tilde{\mathcal{T}}(\delta (\beta -\beta _{0}))=\delta (\beta -\beta
^{\prime }).
\]

Now, consider a doubling procedure of a Hilbert space
\[
\mathcal{H}\rightarrow \mathcal{H}_{A}\otimes \mathcal{H}_{B}
\]%
and a map for density matrices $\hat{\rho}$
\[
\mathcal{B}(\hat{\rho})\in \mathcal{H}_{A}\otimes \mathcal{H}_{B},
\]%
with $\hat{\rho}\in \mathcal{H}$, i.e. a given state for the original
Hilbert space is mapped in a density-like matrix in the doubled space. If
\[
Tr_{A}(\mathcal{B}(\hat{\rho}))=Tr_{B}(\mathcal{B}(\hat{\rho}))=\hat{\rho},
\]%
we say that the $\mathcal{B}(\hat{\rho})$ broadcasts $\hat{\rho}$.

The cloning map, Eq. (\ref{c1}), broadcasts $\hat{\rho}$, but the lack of
linearity implies that superpositions like $\hat{\rho}=\sum_{j\in
Z_{2}}a_{j}\rho _{j}$ cannot be cloned and, as such  are not broadcasted
by the cloning map.

As in the usual no-cloning case, there is a no-broadcasting theorem that
asserts that the above procedure cannot be achieved for an arbitrary
density matrix \cite{barnum}. Indeed, it is not difficult to find
examples of where broadcasting is not achieved. Consider
\begin{eqnarray}
\rho ^{\prime }=\mu \rho \otimes |0\rangle \langle 0|+(1-\mu )|0\rangle
\langle 0|\otimes \rho , \label{1ttt}
\end{eqnarray}
where $\mu \in \lbrack 0,1]$, and $\rho $ is the density operator associated
to the thermofield vacuum, the partial traces are given by $Tr_{A}(\rho
^{\prime })=\mu |0\rangle \langle 0|+(1-\mu )\rho $ and $Tr_{B}(\rho
^{\prime })=\mu \rho +(1-\mu )|0\rangle \langle 0|$ and consequently $\rho
^{\prime }$ do not comes from a broadcasting. On the other hand, other
possible state
\[
\rho ^{\prime \prime }=\mu \rho \otimes \rho +(1-\mu )|0\rangle \langle
0|\otimes |0\rangle \langle 0|,
\]%
has $Tr_{A}(\rho ^{\prime \prime })=Tr_{B}(\rho ^{\prime \prime })=\mu \rho
+(1-\mu )|0\rangle \langle 0|$, and consequently is a candidate to
broadcasted state.

We can introduce a more general notion of broadcasting in the thermofield
context, where although we do not require the complete broadcasting, we
require that a map leads to two different density operators associated
with thermofield vacua at different temperatures. We can write
\begin{eqnarray}
Tr_{A}(\mathcal{B}(\hat{\rho}_{\beta })) &=&\mathcal{T}(\hat{\rho}_{\beta })=%
\hat{\rho}_{\beta ^{\prime }} \\
Tr_{B}(\mathcal{B}(\hat{\rho}_{\beta })) &=&\mathcal{T}^{\prime }(\hat{\rho}%
_{\beta })=\hat{\rho}_{\beta ^{\prime \prime }}
\end{eqnarray}%
where $\mathcal{T}$ and $\mathcal{T}^{\prime }$ are maps in Eq. (\ref{tpo}),
that connect the new density operators associated to thermofield vacua at
temperatures ${\beta ^{\prime }}^{-1}$ and ${\beta ^{\prime \prime }}^{-1},$
respectively. Then the state in Eq. (\ref{1ttt}) is associated with such a
transformation, where the cases $\mu =1$ and $\mu =0$ represent the
extremal, $Tr_{A}(\rho ^{\prime })=\rho $ and $Tr_{B}(\rho ^{\prime
})=|0\rangle \langle 0|$, respectively.

Let us consider the state $\mu \rho +(1-\mu )|0\rangle \langle 0|$ as the
passage of a finite temperature state $\hat{\rho}$ to a zero temperature
state $|0\rangle \langle 0|$, such that the state corresponds to a mixture
state of two different thermal baths. Let us consider a more general case $%
\mu \rho _{\beta }+(1-\mu )\rho _{\beta ^{\prime }}$. Taking the trace, we
have
\[
\mu Tr(\hat{\rho}_{\beta }\hat{\mathcal{O}})+(1-\mu )Tr(\hat{\rho}_{\beta
^{\prime }}\hat{\mathcal{O}}),
\]%
that associates two mean expectation values with the observable $\hat{%
\mathcal{O}}$ for different temperatures, such that
\[
\mu \langle \hat{\mathcal{O}}\rangle _{\beta }+(1-\mu )\langle \hat{\mathcal{%
O}}\rangle _{\beta ^{\prime }}.
\]%
This description is of particular interest in metastable states or
non-equilibrium situation, where the states are connected by means of  two
temperatures near to each other. The corresponding thermofield state
associated to this case is
\[
|0(\beta ,\beta ^{\prime })\rangle =\sqrt{\mu }|0(\beta )\rangle +\sqrt{%
(1-\mu )}|0(\beta ^{\prime })\rangle .
\]%
States at different temperatures are orthonormal, and are written as
\[
\langle 0(\beta ,\beta ^{\prime })|\hat{\mathcal{O}}|0(\beta ,\beta ^{\prime
})\rangle =\mu \langle \hat{\mathcal{O}}\rangle _{\beta }+(1-\mu )\langle
\hat{\mathcal{O}}\rangle _{\beta ^{\prime }}.
\]%
This relation gives us another prospect for the description of states with
different vacua at finite temperature involved. Although in a
non-equilibrium case  a given temperature is not an appropriate parameter,
we can consider superpositions of thermofield vacua at different
temperatures,  in such a way that the expectation value of an observable
with respect to $|0(\beta ,\beta ^{\prime })\rangle $ provides a measurable
estimation. 

\section{Conclusions}

In this paper, we have considered quantum information protocols involving thermofield superpositons of the thermal vacuum and its first excitation 
in the non-tilde sector, a thermofield qubit. We derived a generalized expectation for thermofield qubits and considered the action of gate operations. 
We also proposed QT propocols involving thermofield states. The QT of thermofield qubits was implemented exploring the algebraic properties of 
these states, by means of which we have incorporated the presence of temperature naturally, according to TFD procedure \cite{santanaN}. With 
this approach, we also discussed the case where Alice and Bob are in different temperatures and share a non-local channel of entangled thermofield
states with two different temperatures. Our results show that quantum teleportation can be achieved even if Alice and Bob are in different temperatures. As a particular case for the action of gate operations, we considered Gibbs-like density 
operators under gate actions. By considering the Mandel parameter for thermofield states, we also discussed its changing under gate
operations. We also have discussed no-cloning theorem in TFD \cite{prudencio1} and 
considered the non-broadcasting problem for a thermal context, where we 
we considered maps connecting thermofield vacua, no-broadcasting theorem \cite{barnum} and 
superposition of thermofield vacua and thermofield states at different temperatures. Metastable and non-equilibrium scenarios were also discussed. 

One important aspect to be explored in a more fundamental point is the relation of temperature and non-locality. By exploring this point with 
TFD approach we can circunvent some dificults associated to the inclusion of temperature by indirect means. In fact, 
we considered superpositions of thermofield states in protocols where thermal baths at different temperatures are linked by non-local channels. 
The locality of thermal effects, in the sense that they 
are localized to a given region and cannot be moved by non-local effects is a important point that has to be explored in more detail 
in another place. Some neglected points as dissipative effects affecting dynamically thermofield qubits states will be also explored in somewhere else, 
although some previous proposes have touched this question\cite{tay,khanna2,rakhimov}.

\section{Acknowledgments}

The authors thank CNPq (Brazil) and FAPEMA (Brazil) for financial support of this work partially. TP also thanks Enxoval-project Proquali-PPPG/UFMA.


\end{document}